\documentclass[pdflatex,sn-mathphys]{sn-jnl}
\usepackage{times}
\usepackage{graphicx}
\usepackage{dcolumn}
\usepackage{bm}
\usepackage{mathtools}
\usepackage[english]{babel}
\usepackage{physics}
\usepackage{enumitem}
\usepackage{mathrsfs}
\usepackage{xfrac}
\usepackage{hyperref}
\usepackage{footmisc}
\usepackage{hypcap}
\usepackage{makecell}
\usepackage{url}
\usepackage{ifsym}
\usepackage{bbold}
\usepackage{gensymb}
\usepackage{multirow}
\usepackage{array}
\usepackage{siunitx}
\usepackage{nicefrac}

\jyear{2021}

\begin{document}

\title[Analogue quantum simulation of the Hawking effect in a polariton superfluid]{Analogue quantum simulation of the Hawking effect in a polariton superfluid}

\author*[1]{\fnm{Maxime} \sur{Jacquet}}\email{maxime.jacquet@lkb.upmc.fr}
\equalcont{These authors contributed equally to this work.}
\author[1]{\fnm{Malo} \sur{Joly}}
\equalcont{These authors contributed equally to this work.}
\author[1]{\fnm{Ferdinand} \sur{Claude}}
\author[2]{\fnm{Luca} \sur{Giacomelli}}
\author[1]{\fnm{Quentin} \sur{Glorieux}}
\author[1]{\fnm{Alberto} \sur{Bramati}}
\author[2]{\fnm{Iacopo} \sur{Carusotto}}
\author[1]{\fnm{Elisabeth} \sur{Giacobino}}

\affil*[1]{\orgdiv{Laboratoire Kastler Brossel}, \orgname{ Sorbonne Universit\'{e}, CNRS, ENS-Universit\'{e} PSL, Coll\`{e}ge de France}, \orgaddress{\street{4 Place Jussieu}, \city{Paris}, \postcode{75005}, \country{France}}}

\affil[2]{\orgdiv{INO-CNR BEC Center and Dipartimento di Fisica}, \orgname{Università di Trento}, \orgaddress{\street{via Sommarive 14}, \city{Trento}, \postcode{I-38123}, \country{Italy}}}

\abstract{Quantum effects of fields on curved spacetimes may be studied in the laboratory thanks to quantum fluids.
Here we use a polariton fluid to study the Hawking effect, the correlated emission from the quantum vacuum at the acoustic horizon.
We show how out-of-equilibrium physics affects the dispersion relation, and hence the emission and propagation of correlated waves: the fluid properties on either side of the horizon are critical to observing the Hawking effect.
We find that emission may be optimised by supporting the phase and density of the fluid upstream of the horizon in a regime of optical bistability.
This opens new avenues for the observation of the Hawking effect in out-of-equilibrium systems as well as for the study of new phenomenology of fields on curved spacetimes.}


\maketitle

\section{Introduction}\label{sec:intro}

Analogue gravity is a type of analogue quantum simulation that enables the laboratory study of quantum field theories on curved spacetime via the equivalence between the kinematics of excitation in material systems and of massless fields in astrophysics~\cite{barcelo_analogue_2011,jacquet_next_2020}.
For example, a one-dimensional trans-sonic fluid flow that is a fluid whose flow velocity goes from being sub- to super-sonic forms an acoustic horizon where the flow velocity of the fluid equals the speed of sound~\cite{Unruh}.
In this configuration, excitations of the acoustic field behave as though they propagated on an effectively curved-spacetime whose properties are directly controlled by the geometry of the fluid flow~\cite{visser_acoustic_1998}, thus simulating a Black Hole.
Importantly, quantum fluctuations of the vacuum scatter at the acoustic horizon, causing the emission of correlated waves on both sides of the horizon by the Hawking effect (HE)~\cite{Unruh,Hawking}.

Experimental evidence for correlated emission by the HE was recently reported in analogue gravity setups based on classical~\cite{euve_observation_2016,euve_scattering_2020} and quantum fluids~\cite{munoz_de_nova_observation_2019}.
While the thermal fluctuations of classical fluids overpower quantum fluctuations at the horizon such that vacuum emission cannot be observed there, this can be done with quantum fluids.
Vacuum emission would yield a non-separable state at the output~\cite{campo_inflationary_2006,giovanazzi_entanglement_2011,Busch_entanglementHR_2014,finazzi_entangled_2014,de_nova_violation_2014,nova_entanglement_2015,boiron_quantum_2015,finke_observation_2016,coutant_low-frequency_Vries_2018,coutant_low-frequency_Gennes_2018}, whose degree of entanglement could be quantified from the density and correlation spectra~\cite{jacquet_influence_2020,isoard_bipartite_2021}.

Although most theoretical work on the HE has been dedicated to analogues based on atomic Bose-Einstein condensate (BEC)~\cite{carusotto_numerical_2008,balbinot_nonlocal_2008,Recati_2009,parentani_vacuum_2010,larre_quantum_2012,michel_phonon_2016,wang_correlations_2017,robertson_assessing_2017,fabbri_momentum_2018,isoard_departing_2020}, correlated emission with comparable properties may also be observed in quantum fluids of microcavity polaritons~\cite{Solnyshkov,Gerace,Grissins} where a sonic horizon has already been experimentally realised in one- and two-dimensional flows~\cite{Nguyen,jacquet_polariton_2020}.
In both quantum fluids, the HE manifests itself as the emission of collective Bogoliubov excitations that propagate in opposite directions on either side of the horizon.

The main difference between quantum fluids of atoms and of polaritons is that the latter are intrinsically out of thermal equilibrium.
Radiative and nonradiative dissipative processes in microcavities must be compensated for by optical pumping, and so the non-equilibrium state is not determined by thermodynamic equation conditions~\cite{carusotto_quantum_2013}.
The radiative decay of polaritons does not solely render real-time, in-situ diagnosis of the fluid properties possible (a notable experimental simplification compared with Bose-Einstein condensates of atoms), it also is at the origin of a unique phenomenology in the collective dynamics.
Specifically, in the regime of high fluid density of interest to horizon physics, a gap may open between the dispersion relation and the frequency of the pump~\cite{ciuti_quantum_2005,claude2021highresolution}.
Here we show how this affects the strength of the HE and how, in turn, two-point correlations can be used as a diagnostic for out-of-equilibrium effects.

In this paper, we explore the parameter space of quantum fluids of polaritons and identify a regime that is specifically favourable to the formation of correlations by the HE.
The hydrodynamics of the fluid are controlled by its density and phase profiles, which are in turn connected with the optical bistability of the system (the hysteresis cycle of its polariton-density-to-optical-power relationship)~\cite{baas_bista_2004}, and so we investigate vacuum emission by the HE from this perspective.
We study the influence of the fluid density on either side of the horizon on vacuum emission.
In doing so, we explain how out-of-equilibrium effects in the configurations considered in~\cite{Gerace,Nguyen,Grissins} limit the emission of Bogoliubov excitations, and we show how to engineer the fluid such that the strength and spatial extension of the correlation signal become amenable to experimental detection.
We obtain an order of magnitude increase in both the strength and length of quantum correlations from vacuum fluctuations compared to~\cite{Gerace,Nguyen,Grissins}.
Fine control upon the working point provides us with a better understanding of the influence of the properties of the quantum fluid of polaritons on emission by the HE in systems out of equilibrium.
Our results open the way to the experimental observation of the HE in polaritonic systems.

\section{Polariton fluid and analogue gavity }\label{sec:setup}

\begin{figure}[hb]
    \centering
    \includegraphics[width=.7\textwidth]{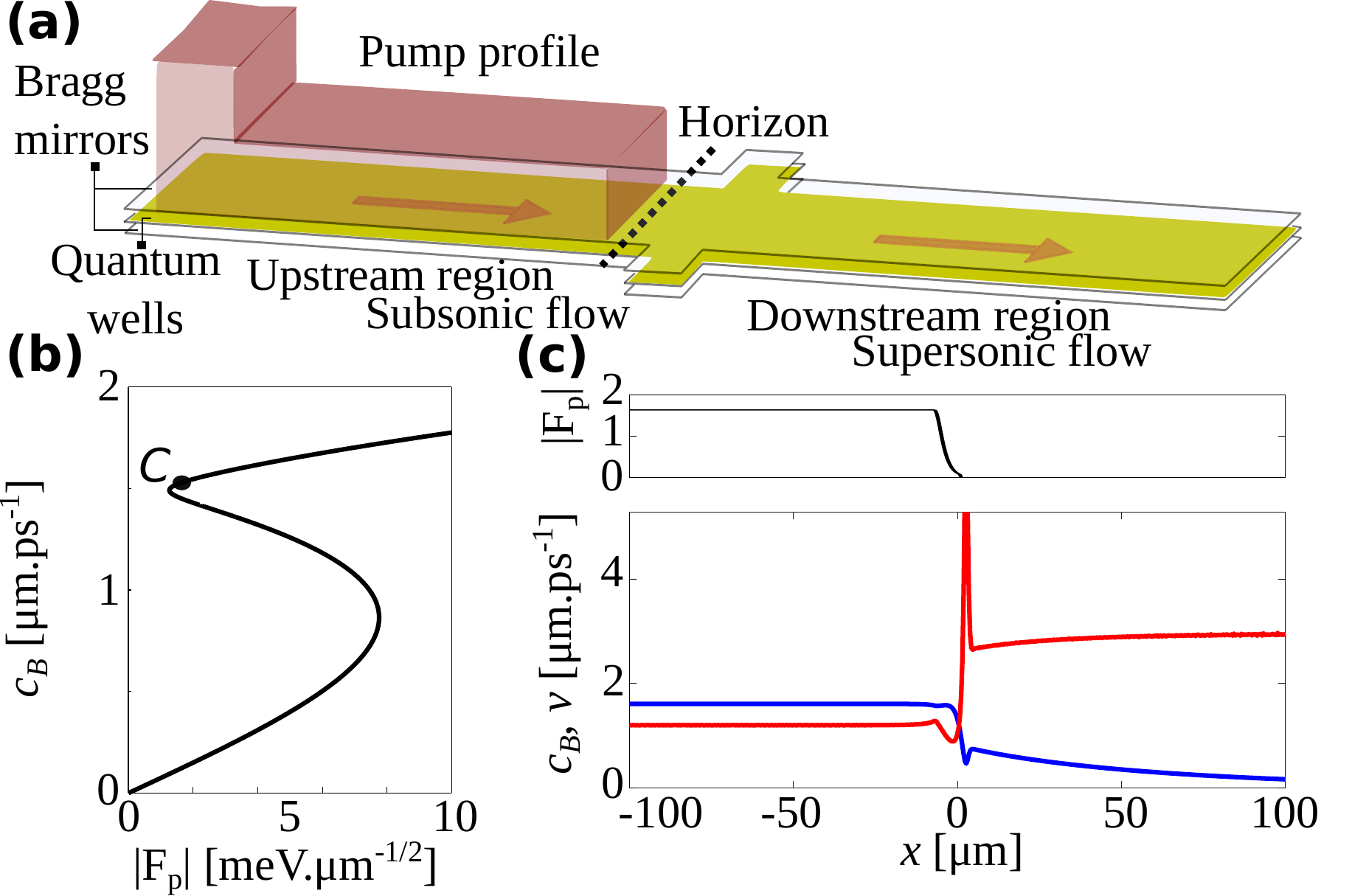}
    \caption{\textbf{Waterfall geometry with a polariton fluid}.
    \textbf{(a)} Sketch of the system: the pump spot is structured in a two-steps profile: on the first step, the fluid is set above the bistable regime ($\lvert F_p\rvert$=9) on the second step the fluid density is supported near the turning point of the bistability loop ($\lvert F_p \rvert$=1.2, point $C$ in \textbf{(c)}). The angle of the pump $k_p$ field creates a transsonic fluid flow across an attractive obstacle at $x=0$.
    \textbf{(b)} Bistability loop.
    \textbf{(c)} Spatial properties of the fluid when setting the fluid density and phase in the upstream region as close as possible to point $C$ of the bistability loop. Black, pump intensity; red, fluid velocity; blue, speed of excitations.}
    \label{fig:system}
\end{figure}

In order to reproduce the kinematics of a scalar quantum field near an event horizon, we use the model of the so-called waterfall geometry illustrated in Fig.~\ref{fig:system} in laboratory frame coordinates $x$ and $t$.
This flow profile is realised in a polariton wire, a laterally confined microcavity in which the polariton dynamics are effectively one-dimensional.
The microcavity is pumped with a continuous wave, coherent pump laser incident at a given angle with respect to the normal to form a stationary flow along the wire.
The light field is structured in a step-like intensity profile (black line in Fig.~\ref{fig:system} \textbf{(c)}).
The first step of high intensity excites the fluid in the nonlinear regime while the second step supports the fluid density at a tunable working point.
As in~\cite{Nguyen,Grissins}, the cavity features an attractive defect (a localised $\SI{1}{\micro\meter}$ long broadening of the wire) placed after the region where the pump lies.

Our system is a one-dimensional quantum fluid of exciton-polaritons whose flow velocity goes from being sub- to super-sonic, thus forming a sonic horizon where the local flow velocity of the fluid equals the local speed of sound.
For now, we begin with the theoretical description of a homogeneous quantum fluid and of the propagation of quantum fluctuations of its phase and density.

Exciton-polaritons are quasi-particles resulting from the interaction of light with matter in a semiconductor microcavity. Photons emitted by a laser are sent into a cavity formed by two Bragg mirrors, wherein their dispersion is the usual Fabry-Perot dispersion, giving them an effective (very low) mass.
These trapped photons create excitons, which are bound electron-hole pairs, in the semiconductor microcavity.
Strong coupling between the photons and excitons trapped in quantum wells gives rise to two eigenstates for the total Hamiltonian, known as the lower polariton (LP) and upper polariton (UP) branches, separated by the Rabi splitting.
Furthermore, the Coulomb interaction between excitons results in an effective non-linearity for exciton-polaritons (polaritons).
The dynamics of the mean-field are governed by a generalised Gross-Piteavskii equation, which leads to Euler and continuity equations describing the system as a quantum fluid.
Historically, polaritons have first been described as two-dimensional quasi-particles~\cite{ciuti_quantum_2005}, although the theory may be reduced to one-dimensional cavities called wires~\cite{Gerace,Nguyen,Grissins}, as in the present case.

In our case, all energies involved are small compared to the Rabi splitting so the exciton-polariton system can be described by the mean field approximation \cite{carusotto_quantum_2013}.
At this level the system is described by a single scalar field $\Psi$, the field of lower polaritons, whose dynamics are governed by the driven-dissipative Gross-Pitaevskii equation (GPE)
\begin{equation}\label{eq:ddGPE}
     i\partial_t \Psi(x,t) =  F_p(x,t)+\left(\omega_0 - \frac{\hbar}{2m^*} \partial_x^2 + V(x) + g\lvert\Psi(x,t)\rvert^2 - i\frac{\gamma}{2}\right)\Psi(x,t).
\end{equation}
$F_p$ is the field of the pump laser, $\omega_0$ is the frequency of the lower polaritons at the bottom of the branch, $m^*$ is their effective mass, $V$ is the `external potential' (that is controlled via the cavity geometry), $g$ is the effective non-linearity, $\gamma$ is the loss rate.
The field $\Psi(x,t)$ is written in the laboratory frame.

In order to make the link between this Gross-Pitaevskii equation and the description of the  polariton ensemble as a fluid, we perform the Madelung transformation: we write the field of lower polaritons as $\Psi(x,t) = \sqrt{n(x,t)} e^{i\theta(x,t)}$ and insert this expression into Eq.~\eqref{eq:ddGPE}.
We take the real and imaginary parts to arrive at the Euler and continuity equations for the polariton fluid~\cite{carusotto_quantum_2013}:
\begin{equation}\label{eq:hydrodynamic}
\begin{split}
    & \partial_t \theta + \frac{m^* v^2}{2\hbar} + \frac{\hbar}{2m^*}\frac{\partial_x^2\sqrt{n}}{\sqrt{n}} + V + gn + \frac{\Re{F_pe^{-i\theta}}}{\sqrt{n}}= 0,\\
    & \partial_t n + \partial_x (n v) = \gamma n - 2\Im{F_pe^{-i\theta}}\sqrt{n}
\end{split}
\end{equation}
with \begin{equation}\label{eq:velocity}
v = \frac{\hbar}{m^*}\partial_x \theta.
\end{equation}
The first and second equations of~\eqref{eq:hydrodynamic} correspond to the Euler and continuity equations of atomic Bose-Einstein condensates (BECs), albeit with terms coming from coherent pumping, losses and quantum pressure.
We see that the properties of the fluid depend on two parameters, namely its density $n=\lvert\Psi\rvert^2$ and phase $\theta$.
The spatial variations of the phase are encapsulated in $v$, which we identify from Eq.~\eqref{eq:velocity} as the flow velocity of the fluid.
Via $n$ and $\theta$, the geometry of the flow may be all-optically controlled by engineering the profile and phase of the pumping laser.

Let us first consider the case of a non-dissipative, non-driven fluid (leave out $\gamma$ and $F_p$).
We also neglect the quantum pressure so Eqs.~\eqref{eq:hydrodynamic} become
\begin{equation}\label{eq:hydrodynamic2}
\begin{split}
    & \partial_t \theta + \frac{m^* v^2}{2\hbar}  + V + gn = 0\\
    & \partial_t n + \partial_x (n v) = 0.
\end{split}
\end{equation}
We investigate small excitations in this system, i.e. low-$k$ phononic modes on top of the fluid (see section~\ref{subsec:fluid} for details).
the linearisation of Eqs.~\eqref{eq:hydrodynamic2} around a background state yields a wave equation equation that is strictly isomorphic to the wave equation of a massless scalar field $\Psi_1$ (the Klein-Gordon equation) on a 1+1D curved spacetime~\footnote{Repeated indices are automatically summed over, following Einstein summation convention.}, $   \Delta\Psi_1\equiv\frac{1}{\sqrt{-\eta}}\partial_\mu\left(\sqrt{-\eta}\eta^{\mu\nu}\partial_\nu\Psi_1\right)=0$, with the metric tensor
\begin{equation}
    \label{eq:metric}
    \eta_{\mu\nu}=\frac{n}{c^2}\begin{pmatrix}
-\left(c^2-v^2\right) & -v\\
-v & 1 
\end{pmatrix},
\end{equation}
$c\propto \sqrt{n}$ the `speed of sound' and $\eta=\mathrm{Det}(\eta_{\mu\nu})$.
The various components of this `acoustic metric'~\cite{Unruh} are given by the fluid velocity.
Notably, there is an event horizon where $v=c$ (the time component of the metric goes to zero)~\cite{visser_acoustic_1998}.
So optically controlling the flow velocity and the speed of sound permits the engineering of an effectively curved spacetime for acoustic waves.
Crucially, this is also valid for the quantized acoustic field, such that correlated pairs are emitted at the horizon by the Hawking effect~\cite{Hawking,carusotto_numerical_2008}.

In paragraph~\ref{subsec:oep} we will show that the driven-dissipative nature of polaritons does not modify the kinematics of excitations.
On the contrary, these additional interesting properties can be harnessed to reveal new effects.

\begin{figure*}[!ht]
    \centering
    \includegraphics[width=\textwidth]{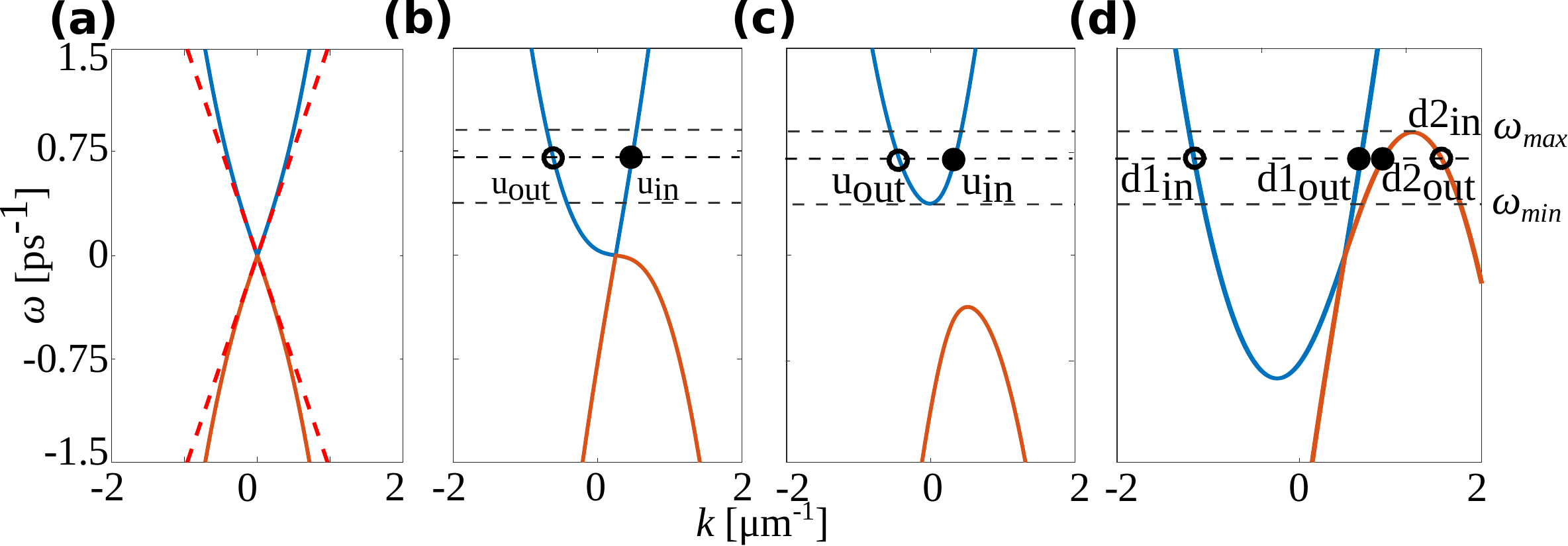}
    \caption{\textbf{Bogoliubov dispersion.}
    \textbf{(a)} Dispersion in the co-moving frame of the fluid at point $C$ in Fig.~\ref{fig:system}. Real part of the dispersion~\eqref{eq:bog_disp_C} for a pump vector $k_p = \SI{0.25}{\micro\meter^{-1}}$. Blue, $\omega_+$, positive-norm modes; orange, $\omega_-$, negative norm modes. Red dashed lines, speed of sound.
    The dispersion in the laboratory frame is obtained by Doppler-shifting the fluid frame dispersion:
    \textbf{(b)} Laboratory-frame subsonic dispersion exactly at point $C$ (Eq.~\eqref{eq:bog_disp_C}).
    \textbf{(c)} Laboratory-frame subsonic dispersion away from point $C$ on the upper branch of the bistability loop (Eq.~\eqref{eq:bog_disp_polariton_lab}).
    \textbf{(d)} Laboratory-frame supersonic dispersion of a ballistic fluid (Eq.~\eqref{eq:disprelbal}).
    Blue, positive-norm branch; orange, negative-norm branch.
    Filled dots, local modes of positive group velocity; circles, local modes of negative group velocity.
    $\omega_{min}$, lower frequency of the gapped positive-norm branch; $\omega_{max}$, upper frequency of the negative-norm branch.}
    \label{fig:disprel}
\end{figure*}
\subsection{Bogoliubov excitations}\label{subsec:fluid}

We now study the propagation of small fluctuations such as quantum fluctuations in this fluid using the linearisation of the Gross-Pitaevskii equation with the Bogoliubov method.

Assuming the pump beam to be monochromatic with frequency $\omega_p$, $F_p(x,t)=F_p(x)e^{-i\omega_pt}$, we write the polariton field as $\Psi(x,t)=\Psi(x) e^{-i\omega_p t}$. In the steady state the Gross-Pitaevskii Equation \eqref{eq:ddGPE}   becomes
\begin{equation}\label{eq:ddGPE1D}
     \left(\omega_0 - \omega_p - \frac{\hbar}{2m^*} \partial_x^2 + V(x) + g\lvert\Psi(x)\rvert^2 - i\frac{\gamma}{2}\right)\Psi(x)
     + F_p(x) = 0,
\end{equation}

We first consider a configuration where the wire is pumped with a spatially homogeneous and monochromatic pump of incident wavevector $k_p$.
The defect creates an attractive potential $V(x)=-\SI{0.85}{\micro\electronvolt}$ with a gaussian distribution centred at $x=0$.
The phase gradient of the fluid is then set by, and equal to, $k_p$ while its density is homogeneous.
The steady-state GPE~\eqref{eq:ddGPE1D} simplifies to
\begin{equation}\label{eq:homddGPE}
    \left(g\lvert\Psi\rvert^2 -\Delta_p - i\frac{\gamma}{2}\right)\Psi + F_p = 0,
\end{equation}
where $\Delta_p$ is the effective detuning defined as the difference between the pump energy and that of lower polaritons,
\begin{equation}
    \Delta_p = \omega_p - \omega_0 - \frac{\hbar k_p^2}{2m^*}.
\end{equation}

Bogoliubov excitations are then mathematically obtained by linearising the GPE~\eqref{eq:ddGPE1D} around a background: $\Psi \rightarrow \Psi + \delta \Psi$, and $\Psi^* \rightarrow \Psi^* + \delta \Psi^*$. The dynamics of the excitations $(\delta \Psi, \delta \Psi^*)$ is given by the Bogoliubov matrix:
\begin{equation}
i\partial_t\mqty(\delta\Psi \\ \delta\Psi^*) = \mathcal{L}\mqty(\delta\Psi \\ \delta\Psi^*).
\end{equation}
We go to the reference frame co-moving with the fluid via a Galilean transform ($x \rightarrow x -vt$).
In the special case of a homogeneous system where the interaction energy matches the detuning, $gn=\omega_p-\omega_0-\frac{\hbar k_p^2}{2m^*}$, the Bogoliubov matrix $\mathcal{L}$ can be written in this frame as
\begin{equation}
    \mathcal{L} = \mqty( gn + \frac{\hbar k^2}{2m^*} - i\gamma/2 & gn\;e^{2ikx} \\
           -gn\;e^{-2ikx} & -gn - \frac{\hbar k^2}{2m^*} - i\gamma/2).
\end{equation}
where $k$ is the wavenumber of Bogoliubov excitations in the co-moving frame.
Upon diagonalization, we retrieve the Bogoliubov dispersion relation in this co-moving frame, which relates $k$ to the frequency of the Bogoliubov excitations $\omega$:
\begin{equation}\label{eq:bog_disp_C}
    \omega^{\Delta_p=gn}_\pm
    = \pm \sqrt{\frac{\hbar k^2}{2m^*}\left(\frac{\hbar k^2}{2m^*} + 2gn\right)} - i\gamma/2.
\end{equation}

Figure~\ref{fig:disprel}~\textbf{(a)} shows the real part of Eq.~\eqref{eq:bog_disp_C}, the dispersion curve, which, in this case where $\Delta_p=gn$, is identical to that of atomic BECs.
There are two branches $\omega^{\Delta_p=gn}_\pm$ of the dispersion, which are symmetrical around the point $\omega=0,\,k=0$.
At low $k$, the dispersion curve has a linear slope: $\omega^{\Delta_p=gn}_\pm \xrightarrow[k\rightarrow 0]{} \pm c_B k$, with $c_B$ is the speed of excitations in the fluid, which is also the speed of sound in the fluid.
\begin{equation}\label{eq:cs}
c_B = c_s = \sqrt{\hbar gn/m^*}
\end{equation}
At large $k$, the dispersion is that of free massive particles, $\omega^{\Delta_p=gn}_\pm \xrightarrow[k\rightarrow \infty]{} \hbar k^2/2m^*$.
There, $\lvert\pdv*{\omega^{\Delta_p=gn}_\pm}{k}\rvert > c_s$ the gradient of the dispersion curve is larger than the speed of sound.

\subsection{Acoustic horizon}

The polariton ensemble behaves as a fluid whose dispersive properties depend on its density and velocity, and we will use these properties to study the waterfall geometry, which is the basis of our black hole analogue model.
The waterfall separates two regions of fluid density $n$ and phase $\theta$.
The polariton fluid flows across a defect close to $x=0$ in the positive $x$ direction at velocity $v = \frac{\hbar}{m^*}\partial_x \theta$, as shown in red in Fig.~\ref{fig:system} \textbf{(c)}.
The speed of excitations in the fluid depends on their wavenumber. As mentioned above,  Bogoliubov excitations have a group velocity $\nicefrac{\partial\omega}{\partial k}$.
However, the relevant case for our model is when the excitations have a small wavenumber $k$ and their speed, given by Eq.~\eqref{eq:cs}, does not depend on $k$ and is proportional to the square root of the fluid density (in blue in Fig.~\ref{fig:system} \textbf{(c)}).

As can be seen in Fig.~\ref{fig:system} \textbf{(c)}, the speeds of the fluid and of the excitations, calculated from the GPE equation Eq.~\eqref{eq:ddGPE}, vary a lot when the flow goes across the defect. Upstream of the defect, for $x<\SI{-2}{\micro\meter}$ we have $v<c_B$, the flow is subsonic while downstream of the defect, for $x>\SI{-2}{\micro\meter}$, $v>c_B$, the flow is supersonic, so there is an \textit{acoustic horizon} at $x_H=\SI{-2}{\micro\meter}$. The region where the flow is subsonic is \textit{outside} the horizon, the region where the flow is supersonic is \textit{inside} the horizon \footnote{There exist stricter definitions of the curvature of the effective spacetime in analogue gravity, see~\cite{jacquet_influence_2020}, but the related considerations do not impact the conclusions we draw in the present work.
Strictly speaking, the interface at $x=\SI{-2}{\micro\meter}$ is a sonic horizon only at frequencies for which there are two propagating local modes in the upstream region and four propagating local modes (including negative-norm modes) in the downstream region~\cite{macher_black/white_2009,larre_quantum_2012,Isoard}: that is for $\omega\in[\omega_{min},\omega_{max}]$.}.

In the next section, we will focus on the properties of the fluid in the waterfall configuration and the kinematics of collective (or Bogoliubov) excitations therein.
As can be seen in Fig.~\ref{fig:system} \textbf{(c)}, the flow velocity $v$ is rather flat, except in the near-horizon region.
In particular, $v$ spikes around $x=0$ because of approximate conservation of the flow current there (under dissipation).
Beside this narrow feature, the flow velocity may be treated as homogeneous on either side of the region around $x=0$.
Meanwhile, $c_B$ is flat in the pumped region, slightly bumps between the edge of the pump and the defect as well as just after the defect (under the approximate conservation of the flow current mentioned above) and then decreases as the polariton population decays.
For the sake of simplicity, we may consider that the speed of excitations is homogeneous on either side of the region around $x=0$.

We will first discuss in more detail the dispersive properties of the fluid as a function of the parameters of the polariton field and identify the plane waves of this field before explaining how to construct ``global modes'' of the inhomogeneous system (including the horizon). These are the modes in which correlated emission from vacuum occurs.
We then show the influence of out-of-equilibrium physics on the kinematics of Bogoliubov excitations on either side of the horizon.

\subsection{Dispersion relations and optical bistability}

In this section, we treat the system as composed of two regions homogeneous in their properties on either side of the horizon at $x_H=\SI{-2}{\micro\meter}$. 
The properties of the polariton field in the upstream region are set by those of the pump, and they are given by Eq.~\eqref{eq:ddGPE1D}, with no external potential ($V(x)=0$).
By linearizing Eq.~\eqref{eq:ddGPE1D} in these conditions, we obtain the dispersion relation in the laboratory-frame:
\begin{equation}\label{eq:bog_disp_polariton_lab}
    \omega_{\pm} (k) =
    \pm\sqrt{\left(\frac{\hbar \delta k_p^2}{2m^*} - \Delta_p + 2 gn\right)^2 - (gn)^2} + v \delta k_p - i\gamma/2
\end{equation}
with $\Delta_p \coloneqq \omega_p - \omega_0 - \nicefrac{\hbar k_p^2}{2m^*}$, the effective detuning between the pump energy $\hbar\omega_p$ and that of polaritons flow.
$k_p$ is the wave-number of the pump field, $\delta k_p=k-k_p$, and $g$ the interaction strength. The term $v\delta k_p$ is due to the Doppler effect in the laboratory frame \cite{ciuti_quantum_2005}. Note that unlike the configuration considered in the previous section, we do not assume that the interaction energy matches the effective detuning, and the dispersion curve is thus modified.

On the other hand, in the downstream region, the pump field is zero,  so the polariton fluid propagates ballistically there~\cite{amelio_perspectives_2020} and the dispersion is
\begin{equation}
    \label{eq:disprelbal}
    \omega_\pm(k) =
    \pm \sqrt{\frac{\hbar \delta k_0^2}{2m^*}\left(\frac{\hbar \delta k_0^2}{2m^*} + 2gn\right)}
    + v\delta k_0 - i \gamma/2,
\end{equation}
with $\delta k_0=k-k_0$ where $k_0 = \nicefrac{m^* v}{\hbar}$ denotes the wave-number of the ballistic fluid.

In the upstream region, due to presence of the pump laser, the relationship between the density of polaritons $n$ and the intensity of the pump laser $\lvert F_p\rvert^2$ is obtained from squaring Eq.~\eqref{eq:homddGPE}
\begin{equation}
\left(\left(gn - \Delta_p\right)^2 + \frac{\gamma^2}{4}\right) n = \lvert F_p\rvert^2
\end{equation}
In the case where the energy of the laser is above that of the lower polaritons, $\Delta_p > \gamma\sqrt{3}/2$, this equation has several  solutions in $n$ for a given value of $F_p$~\cite{carusotto_parametric_threshold_2005}. This degeneracy of fluid densities is due to optical bistability~\cite{baas_bista_2004}. As shown in Fig.~\ref{fig:system} \textbf{(b)}, there is a hysteresis relationship between the fluid density $n$ and the pump strength $\lvert F_p\rvert$, which comes from the bistability loop (cf Appendix~\ref{app:homogeneous}). 

The previous equation can be written in terms of the velocity of excitations $c_B$.
\begin{equation}
        \left(\left(\frac{m^* c_B^2}{\hbar} - \Delta_p\right)^2 + \frac{\gamma^2}{4}\right) \frac{m^* c_B^2}{g\hbar} = \lvert F_p\rvert^2.
\end{equation}
This shows that bistability has a critical influence on the  propagation of excitations of the fluid, including Bogoliubov excitations.

Under optical bistability, the shape of the dispersion curve (the real part of the dispersion relation~\eqref{eq:bog_disp_polariton_lab}) depends on the working point along the loop, i.e. on $\Delta_p$.
In the special case where $\Delta_p = gn$ (point $C$ in Fig.~\ref{fig:system} \textbf{(b)}), one recovers the behaviour given in Eq.~\eqref{eq:bog_disp_C} and the dispersion curve has a linear slope at low $\delta k_p$ where interactions are phononic, while the dispersion at high $k$ is that of free massive particles, $\omega^{\Delta_p=gn}_\pm \xrightarrow[k\rightarrow \infty]{} \hbar k^2/2m$~\footnote{
There, $\lvert\pdv*{\omega^{\Delta_p=gn}_\pm}{k}\rvert > c_{B\lvert_{\omega=0}}$ --- the gradient of the dispersion curve is larger than the speed of sound, so the dispersion is said to be `superluminal' (in analogy with superluminal corrections to the dispersion in eg~\cite{brout_hawking_1995,corley_hawking_1996}). This superluminal correction to the dispersion allows for the propagation of modes against the flow velocity of the fluid inside the horizon. 
}.
Because of the linear, sound-like, dispersion of excitations at short $k$, point $C$ is referred to as the ``sonic point'' of the bistability loop.

Operation at $\Delta_p < gn$ is also possible, in which case the linear behaviour at short $k$ disappears and the dispersion becomes quadratic.
As we will show in Section~\ref{subsec:oep}, this bears consequences on the emission and propagation of Bogoliubov excitations in the fluid.
For now, we consider that $\Delta_p = gn$.
In the downstream region where the fluid is ballistic, interactions at low $\delta k_0$ are phononic, as shown in Eq.~\eqref{eq:disprelbal}.

\subsection{Modes of the system \label{subsec:modes}}

In the configuration of Fig.~\ref{fig:system} \textbf{(a)}, the fluid flow is transsonic: it goes from being sub- to supersonic with a sonic horizon ($v=c_B$) at $x=\SI{-2}{\micro\meter}$. 
We plot the dispersion curve in the laboratory frame (the real part of Eq.~\eqref{eq:bog_disp_polariton_lab}) of the subsonic fluid flow upstream in Fig.~\ref{fig:disprel} \textbf{(b)}/\textbf{(c)}, and of the supersonic fluid flow downstream in Fig.~\ref{fig:disprel} \textbf{(d)}. Blue (orange) curves correspond to $\omega_+$ ($\omega_-$) solutions of Eq.~\eqref{eq:bog_disp_polariton_lab}. In the subsonic case of Fig.~\ref{fig:disprel} \textbf{(b)}, the sonic behaviour close to $\omega=0$, with a slope shifted by the Doppler effect, can easily be seen. In Fig.~\ref{fig:disprel} \textbf{(d)} the fluid is supersonic and the shape of the dispersion curve of the excitations in the laboratory frame is changed a lot. The sonic behaviour at $\omega=0$ manifests itself by a discontinuity in the slope of the dispersion curve~\eqref{eq:disprelbal} in Fig.~\ref{fig:disprel} \textbf{(d)}.

Note that in the rest frame of the fluid, the $\omega_+$ ($\omega_-$) modes, which are the positive (negative) norm modes ~\cite{castin_lecture_notes} have positive (negative) energies.
However, in the laboratory frame, the Doppler effect modifies the shape of the dispersion relation.
For subsonic fluid flows, the $\omega_-$ branch is at negative laboratory frame energies.
For supersonic flows, part of the negative norm  $\omega_-$ branch is pulled up to positive laboratory frame energies (up to a maximum energy which we denote by $\omega_{max}$) while part of the positive norm  $\omega_+$ branch is pulled down to negative laboratory frame energies. This has a critical effect on the transmission of the excitations at the horizon.

Now that we have described the dispersive properties of the transsonic fluid, we consider the kinematics of Bogoliubov excitations therein.
Because of the steady-state condition of the system, these are plane wave modes.
Eq.~\eqref{eq:bog_disp_polariton_lab} is a fourth-order polynomial, so there are four (positive laboratory-frame frequency) solutions to the equations of motion in each spatial region on either side of the interface.
These solutions are found at the intersection point of line of constant $\omega$ with the dispersion branches at positive energies in Fig.~\ref{fig:disprel} \textbf{(c)} and \textbf{(d)}.
Although these solutions share the same $\omega$ (which manifests energy conservation in the laboratory frame), they have distinct $k$, i.e. they are local modes of the homogeneous system.
For $\omega>0$ in the upstream region there are two propagating modes of positive norm and two modes of complex $\omega$ and $k$, which are exponentially growing and decaying modes.
For $\omega<\omega_{max}$ in the downstream region, there are four propagating modes, two of which have positive norm while the other two have negative norm.
For $\omega>\omega_{max}$, there are two propagating modes of positive norm and two exponentially growing and decaying modes.

Local modes in a homogeneous region may be sorted by their respective group velocity $v_g = \pdv*{\omega_\pm}{k}$: those which have positive group velocity propagate rightwards (towards positive $x$) while those which have negative group velocity propagate leftwards.
In the supersonic region downstream, because of the superluminal  behaviour of the dispersion relation for large $k$, there are leftward propagating modes. 
This is a specificity of analogue systems based on quantum fluids (be they atomic or polaritonic~\cite{macher_black/white_2009,Recati_2009,corley_hawking_1996}).

From the local modes, one may construct modes of the whole transsonic fluid --- the global modes (GMs)~\cite{castin_lecture_notes}.
These are solutions to the equation of motion that are valid in both regions on either sides of the interface.
GMs correspond to waves scattering at the interface, and they describe the conversion of an incoming field to scattered fields in both regions.
Although the system is driven-dissipative, GMs may be constructed similarly to those of conservative systems, i.e. as superpositions of the plane wave solutions in the two homogeneous regions on either side of the interface~\cite{macher_black/white_2009}.
GMs are identified via their defining local mode depending on its group velocity $v_g$: an \textit{in} GM describes the scattering of an incoming plane wave to various outgoing plane waves while an \textit{out} GM describes a single outgoing plane wave resulting from the scattering of various incoming waves.

The horizon separates two regions of differing properties, so the vector bases of the \textit{in} and \textit{out} GMs are different.
As a result, scattering at the horizon mixes the annihilation and creation operators of the field (as described by a Bogoliubov matrix~\cite{jacquet_analytical_2020,isoard_bipartite_2021}).
The matrix giving the scattering of the operators of the in-going modes $a$ into the operators of the outgoing modes $b$ can be expressed as \cite{Recati_2009}
\begin{equation}
\mqty(b_u \\ b_{d1} \\ b^{\dag}_{d2}) =  
\textbf{S}
\mqty(a_u \\ a_{d1} \\ a^{\dag}_{d2})
\end{equation}
where the indices $u$ and $d$ indicates the upstream and downstream positions. The $\textbf{S}$ matrix elements are determined from the transmission/reflection coefficients of an in-going mode into an outgoing mode.  Because of their negative norm, the $d2$ modes must be quantized using creation operators $a^{\dag} _{d2}$ and $b^{\dag} _{d2}$ rather than annihilation operators. This is at the origin of the Hawking radiation.  
As a result quantum fluctuations in the \textit{in} GMs $\{u_{in},d1_{in},d2_{in}\}$ will be converted into pairs of real excitations in the \textit{out} GMs $\{u_{out},d1_{out},d2_{out}\}$: the Hawking radiation  will occur in correlated pairs $u_{out}-d2_{out}$ (Hawking-partner) on top of the classical background formed by the mean-field of the polariton fluid~\footnote{As in all quantum fluids, there are also quantum correlations between the partner and the companion waves ($d1_{out}-d2_{out}$) and classical correlations between Hawking and the companion wave ($u_{out}-d1_{out}$)~\cite{isoard_bipartite_2021}}.

\subsection{Effects of out-of-equilibrium physics\label{subsec:oep}}
So far we have discussed the dispersive properties of the fluid when $\Delta_p=gn$, that is, when operating at the sonic point $C$ of the bistability loop.
In the regime $\Delta_p<gn$, the microcavity acts as an optical limiter~\cite{ciuti_quantum_2005}: as can be seen in Fig.~\ref{fig:system} \textbf{(b)}, the growth of the speed of excitations $c_B$ with the pump strength on the upper branch of the loop is sub-linear.
While the fluid is stable in this regime, a gap opens between the $\omega_\pm$ branches of the dispersion curve, see Fig.~\ref{fig:disprel} \textbf{(c)}.
We mark the bottom of the $\omega_+$ curve as $\omega_{min}$.

This behaviour is markedly different from that observed in systems like quantum fluids of atoms.
There, the oscillation frequency of the condensate wavefunction corresponds to the chemical potential.
Instead here it corresponds to $\omega_p$. The opening of the gap illustrates how tuning $\Delta_p$ gives access to a unique phenomenology of collective dynamics.
Specifically, the linear behaviour at short $k$ disappears as soon as the gap opens and the dispersion is always quadratic, meaning that the polariton ensemble cannot be superfluid.
As we will see in Section~\ref{subsec:fluconf}, this departure from superfluid propagation modifies the density of the fluid in the region $x<0$ as a function of the pump strength and profile.

In Section~\ref{subsec:modes}, we have seen that the Hawking Effect consists in the mixing of \textit{in} GMs of opposite sign of norm at the horizon, $u_{in}$ from $x<0$ and $d1_{in}$ and $d2_{in}$ from $x>0$.
The downstream modes only exist over the limited interval $0<\omega<\omega_{max}$.
When the gap between $\omega_-$ and $\omega_+$ opens, $u_{in}$ only exists for $\omega>\omega_{min}>0$ so the frequency interval for scattering is reduced to $\omega_{min}<\omega<\omega_{max}$.
We will show in the simulations (cf Section~\ref{subsec:corrspec}) that the strength of vacuum emission by the HE is thus decreased.

In brief, when $\Delta_p<gn$, out-of-equilibrium physics manifests itself in the opening of a gap between the branches of the dispersion relation and a modification of the shape of the dispersion to a purely quadratic form.
This affects the generation of Bogoliubov excitations as well as their propagation in the fluid.

\section{Emission by the Hawking effect}\label{sec:Hawking}
We now perform calculations with the cavity parameters of \cite{Gerace}: $\hbar\gamma =$ \SI{0.047}{meV}, $\hbar g =$ \SI{0.0003}{meV \micro\meter}, $m^* = 3 \cdot 10^{-5} m_e$.
Importantly, $\omega_p-\omega_0 =$ \SI{0.49}{meV} was kept constant throughout.

We study vacuum emission via non-local correlations in the fluid density~\cite{carusotto_numerical_2008}, which we quantify with the normalised spatial correlation function
\begin{equation}
\label{eq:g2}
    g^{(2)}(x,x') = \frac{G^{(2)}(x,x')}{<n(x)><n(x')>}.
\end{equation}
$G^{(2)}(x,x')$ is the two-point correlation function of the field (cf Appendix~\ref{app:twa}).

\begin{figure*}[!ht]
    \centering
    \includegraphics[width=\textwidth]{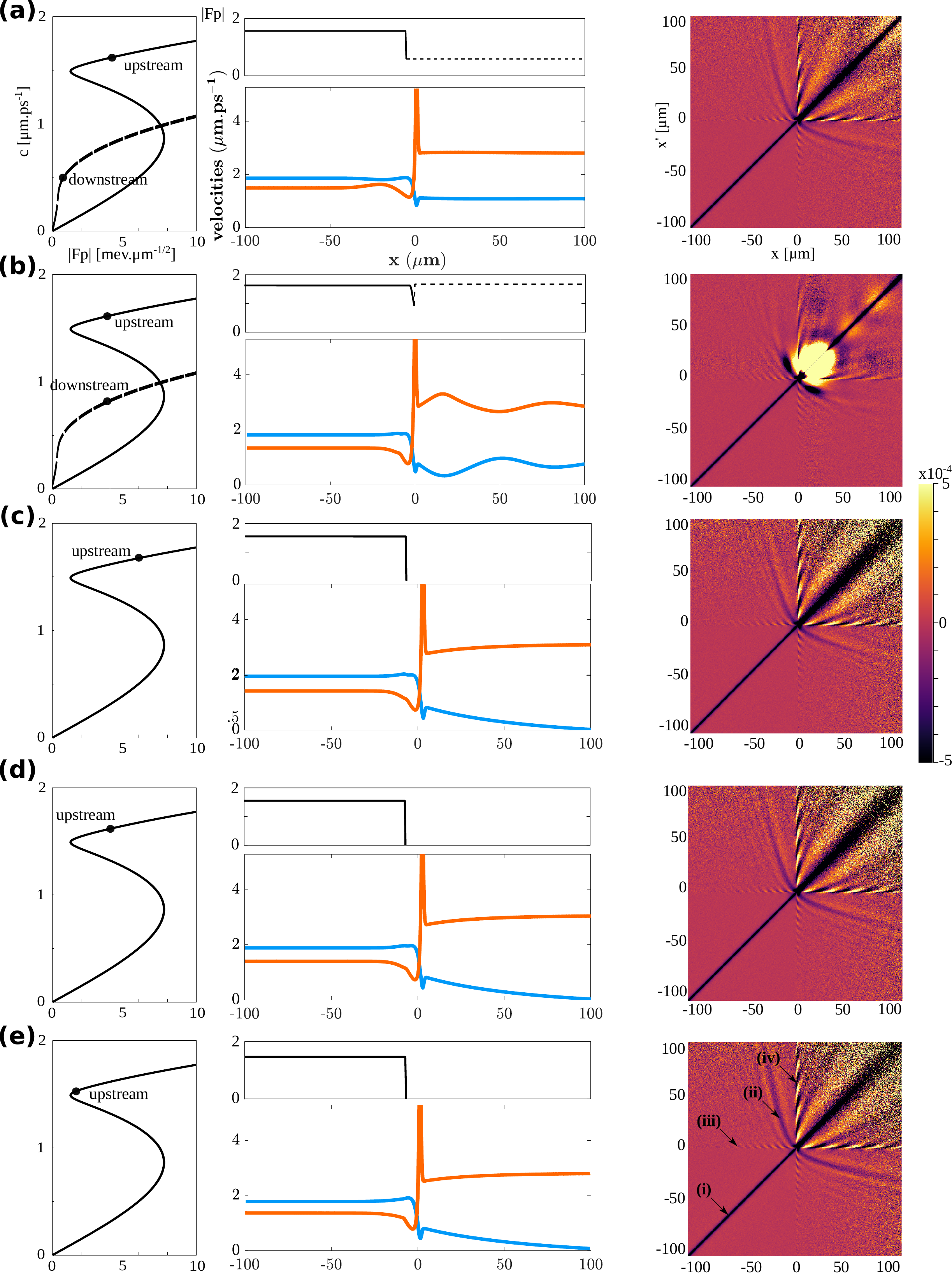}
    \caption{\textbf{Correlated emission as a function of the trans-sonic flow profile.}
\textbf{Left column}, bistability loop: solid black, upstream region; dashed grey, downstream region.
\textbf{Middle column}, pump strength $\lvert F_p\rvert \, [\mathrm{ps}^{-1}.$\textmu$\mathrm{m}^{-1/2}]$:  solid black, upstream; dashed black, downstream.
Velocities [\SI{}{\micro\meter\per\pico\second}]. blue, speed of excitations $c_B$; orange, fluid flow velocity $v$ set by $k_{p,u}$. If the pump strength is zero in the downstream region (\textbf{(c)}, \textbf{(d)} and \textbf{(e)}), the fluid propagates ballistically (free evolution with no support) there.
\textbf{Right column}, spatial correlation function $g^{(2)}(x,x')-1$ (Eq.~\eqref{eq:g2}), colour scale from $-1.25\times10^{-4}$ to $1.25\times10^{-4}$.}
    \label{fig:corr}
\end{figure*}

In Fig.~\ref{fig:corr} we plot the operation point on the bistability loop of the fluid on either side of the horizon (solid line, upstream, dashed line, downstream), the pump profile (black line) and ensuing properties of the inhomogeneous fluid -- characterised by its velocity (red line) and the speed of excitations $c_B$ (blue line) -- as well as the resulting density-density correlations~\eqref{eq:g2}.
We are interested in the fluid properties and their influence on correlated emission.
Note that we plot $c_B$ for better comparison between the working point along the bistability loop and the local properties of the inhomogeneous fluid.

\subsection{Fluid configurations\label{subsec:fluconf}}
We consider various flow profiles on either side of the horizon.
The bistability of the fluid on either side of the horizon may be tuned by controlling the wave-number of the fluid in either region by means of the pump ($k_{p,u}$ or $k_{p,d}$ in the up- or downstream region, respectively), see Appendix~\ref{subsec:bista}.
The fluid density may be supported on the higher branch of the bistability loop by means of the pump intensity.
In order to explore the full parameter space we have computed $36$ correlation spectra.
Not all combinations are interesting, though, so in Fig.~\ref{fig:corr} we present five configurations that give typical behaviours: row \textbf{(a)}, the fluid density is set near (but not at) the sonic point in both regions with a jump in $k_p$ at the interface ($k_{p,u} = \SI{0.25}{\per\micro\meter}$, $k_{p,d} = \SI{0.55}{\per\micro\meter}$), the pump strength is set to 0 at $x=-\SI{7}{\micro\meter}$ the horizon; row \textbf{(b)} the fluid density is set near to and away from the sonic point on the upper branch of the bistability loop in the up- and downstream region, respectively, with a jump in $k_p$ at the interface ($k_{p,u} = \SI{0.25}{\per\micro\meter}$, $k_{p,d} = \SI{0.58}{\per\micro\meter}$); rows \textbf{(c)}, \textbf{(d)} and \textbf{(e)}, the fluid density is set gradually closer to the sonic point on the upper branch of the bistability loop in the upstream region ($k_{p,u} = \SI{0.25}{\per\micro\meter}$), the pump strength is set abruptly to zero at $x=-\SI{7}{\micro\meter}$ and so the fluid is left free to evolve from that point on (across the defect into the downstream region).
In configuration \textbf{(e)} the fluid density is supported as close as possible to the sonic point.

In all configurations, the fluid builds up in the region $-\SI{10}{\micro\meter}<x<x_d$: a relatively small amplitude bump in the density forms before the defect (the fluid velocity slightly dips in that region).
On the other hand, while the density of the fluid is mostly flat in the configuration of Fig.~\ref{fig:corr} \textbf{(a)}, in that of Fig.~\ref{fig:corr} \textbf{(b)} its amplitude undulates widely over $\SI{100}{\micro\meter}$ downstream of the horizon before flattening down.
This illustrates how attempting to force the fluid properties to a working point away from the sonic point after it has propagated across an obstacle destabilises it.
Meanwhile, the fluid density decreases exponentially (with a high, ballistic, wavenumber) in configurations \textbf{(c)}, \textbf{(d)} and \textbf{(e)} where there is no pump in the downstream region.

Given the variety of fluid properties and the possible fast variations within, the description of the system as two homogeneous media adopted in Section~\ref{sec:setup} and amenable to analytical solutions is not valid everywhere.
Instead we must calculate the flow profile and quantum fluctuations at all points.
To this end, we use the Truncated Wigner Approximation (see Appendix~\ref{app:twa}) to evolve the wave function and obtain the properties of the fluid at all points in the cavity as well as the dynamics of the Bogoliubov excitations therein.
This numerical method is adapted to analogue systems based on atomic as well as polaritonic quantum fluids~\cite{carusotto_numerical_2008,carusotto_parametric_threshold_2005}.
Here, it enables the study of vacuum emission on highly varying backgrounds.
All maps result from a statistical average over $10^6$ Monte-Carlo realisations.

\subsection{Correlation diagrams\label{subsec:corrspec}}
In all configurations, correlations may be sorted by the spatial region in which the involved modes propagate, which correspond to four quadrants in the plots.
The South West quadrant ($x<0$, $x'<0$) corresponds to correlations in the upstream region; the South East and North West quadrants correspond to correlations across the horizon in the up- and downstream regions; the North East quadrant corresponds to correlations in the downstream region.
All configurations have some common traces, which are most visible in Fig.~\ref{fig:corr} \textbf{(e)}: (i) anti-correlations along the $x=x'$ diagonal that indicate anti-bunching under repulsive polariton interactions; (ii) a negative moustache-shaped trace in the \emph{upstream-downstream} region that indicates correlations across the horizon between Hawking radiation and modes in the downstream region; (iii) oblique interference fringes localised along the $x=0$, $x'>0$ half line (and, symmetrically, the $x'=0$, $x>0$ half line), and another series (iv) of fringes localised along the $x=0$, $x'<0$ half line.
While traces (i) and (ii) are generic features of the HE in quantum fluids, see eg~\cite{carusotto_numerical_2008,Recati_2009,Gerace,Grissins,isoard_departing_2020}, the fringes (iii) and (iv) are new.
They indicate correlations between the propagating modes $u_{out}$ and $d1_{out}$ and a mode bound to the horizon.
This coupling manifests the excitation of quasi-normal modes of the acoustic field under vacuum-driven perturbations~\cite{jolyInterplayHawkingEffect2021}.
(Note that this was not seen in~\cite{Gerace}, where a repulsive defect was used, see Appendix~\ref{app:repdef})

Configuration~\ref{fig:corr} \textbf{(e)} leads to a longuer Hawking moustache (ii) than obtained in other models for a quantum fluid: about \SI{35}{\micro\meter}- and \SI{105}{\micro\meter}-long in the up- and downstream regions, respectively.
It also features stronger correlations than obtained in previous polariton analogue studies, with a maximal relative amplitude of $1.5\cdot10^{-4}$.
In comparison, the Hawking moustache is shorter and weaker in all other configurations: we observe the influence of the regime of density of the fluid properties (working point on the bistability loop on either side of the horizon) on vacuum emission at the horizon and propagation in either region thereafter.
We remark that supporting the density of the fluid in the downstream region as in configuration \textbf{(a)} does not aid emission while coming at a higher technical cost (due to the precise matching of the pump wavenumber across the defect).
Finally, although emission occurs in all configurations, as a comparison between configurations \textbf{(c)}, \textbf{(d)} and \textbf{(e)} shows, its strength is limited by the opening of the gap in the dispersion upstream, which reduces the interval $\{\omega_{min},\omega_{max}\}$ over which the HE occurs.
This is why the emission strength in~\cite{Gerace} was lower than in Fig.~\ref{fig:corr}~\textbf{(e)} although a similar pump profile was used.

In brief, we have established that operating such that the fluid is as close as possible to the turning point $C$ of the bistability loop upstream of the horizon and letting the fluid propagate ballistically downstream enhances the emission and propagation of Bogoliubov excitations.
In that regard, operating with a flat pump profile whose spatial extension is well controlled is better than with a Gaussian profile.

\section{Discussion}
We showed how engineering the density of a quantum fluid of polaritons can enhance the emission and propagation of paired Bogoliubov excitations in a transsonic flow.
Our work sheds light on the interplay between optical bistability and parametric amplification in fluids of light.
The bistable behaviour of a system can thus be exploited to study field theoretic effects like the HE in the laboratory.
Specifically, we have found that fine control over the fluid properties may be achieved with a step-like pump profile.

Here we observed the generation and propagation of paired Bogoliubov excitations of the quantum fluid on either side of a sonic horizon when supporting the density of the fluid at various points in the bistable regime.
Support of an inhomogeneous fluid density and velocity may be achieved by changing the wave-number of the pump.
In an experiment, this is easily implemented with high spatial resolution (limited by diffraction) thanks to spatial light modulators~\cite{vocke_rotating_2018,maitre_dark_soliton_2020,lerario_vortex-stream_2020,lerario_parallel_2020,claude_taming_2020}.
We found that letting the fluid flow ballistically across an attractive defect so as to form a horizon yields Hawking correlations of the order of $10^{-3}$ over more than \SI{100}{\micro\meter}.
Thus we obtained a strong increase of the quantum correlations by the HE: these are a fourfold enhancement in the strength and tenfold enhancement in the propagation length of correlations compared with previous results in quantum fluids of light.
Furthermore, we showed how moving away from this optimal configuration reduces the signal strength and length --- two effects directly linked to the kinematics of Bogoliubov excitations in the out-of-equilibrium fluid.
In this way, our work demonstrates that the correlation traces are a diagnostic for the influence of out-of-equilibrium physics on mode conversion in inhomogeneous flows.
For example, in the paper~\cite{jolyInterplayHawkingEffect2021}, we have explained how a dissipative quench of a mode bound to the horizon yields novel, local correlation traces between a local and propagating modes, i.e. a quasi-normal mode of the acoustic field.

Given the strength of the Hawking correlations reported in this paper, these should be amenable to experimental detection with state-of-the-art apparatus.
In fact, the coupling constant $g$ of experiment~\cite{Nguyen} should yield a 4-fold increase on the correlation amplitude obtained here.
As such, our methods open the way to the theoretical and experimental study of the quantum statistics of the HE in driven-dissipative systems: for example, one could calculate (and observe) the Hawking correlations in reciprocal space~\cite{fabbri_momentum_2018}, thus gaining frequency-resolved information on them~\cite{isoard_departing_2020}, which could in turn be used to measure entanglement in the Hawking emission~\cite{jacquet_influence_2020,isoard_bipartite_2021}.

\bmhead{Acknowledgments}
We thank Michiel Wouters for discussions on dispersion in bistable fluids, Tangui Aladjidi for help with code speed-ups as well as computer power, and Yuhao Liu for his work early in the project.

\section*{Declarations}
\begin{itemize}
    \item Funding: We acknowledge financial support from the H2020-FETFLAG-2018-2020 project ``PhoQuS'' (n.820392). IC and LG acknowledges financial support from the Provincia Autonoma di
Trento and from the Q@TN initiative. QG and AB are members of the Institut Universitaire de France.
    \item Availability of data and materials: The datasets generated during and/or analysed during the current study are available from the corresponding author on reasonable request.
    \item Code availability: The Julia code written during the current study is available from the corresponding author on reasonable request. 
    \item Authors' contributions: MJJ conceived the project. MJ, MJJ and LG carried out the numerical simulations. All authors contributed to the analysis. MJJ, MJ, IC and EG wrote the manuscript.
\end{itemize}

\begin{appendices}
\section{The physical system}\label{app:homogeneous}
In this appendix we present the field theory of polaritons and their fundamental excitations.

Our system is a one-dimensional quantum fluid of exciton-polaritons whose flow velocity goes from being sub- to super-sonic, thus forming a sonic horizon where the local flow velocity of the fluid equals the local speed of sound.
For simplicity, we may consider that the horizon separates two spatial regions whose properties are independent of space --- two homogenous regions, although we shall eventually depart from this simplified picture.
For now, we begin with the theoretical description of a homogeneous quantum fluid and of the propagation of quantum fluctuations of its phase and density.

In the majority of cases the interaction energy does not match the effective detuning and the dispersion curve is thus modified.
Furthermore, writing the density of the fluid as a function of the intensity of the laser yields several solutions~\cite{carusotto_parametric_threshold_2005}.
This degeneracy of fluid densities is due to optical bistability~\cite{baas_bista_2004}, which, as we will show, has tremendous influence on the emission and propagation of excitations of the fluid, including Bogoliubov excitations.
Here we describe the influence of the bistability on the Bogoliubov dispersion.

\begin{figure}[t]
    \centering
    \includegraphics[width=.5\columnwidth]{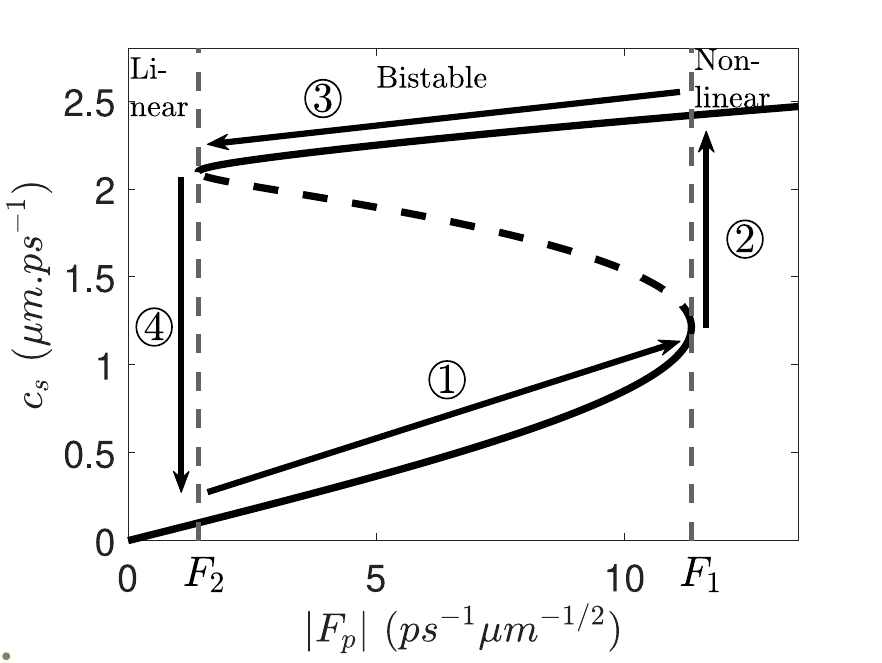}
    \caption{\textbf{Bistability loop for an homogeneous polaritonic fluid.} $\omega_p - \omega_0>0$ and $k_p = 0$. Black, stable points; dashed, unstable points.
    The system is bistable for $F_2<\lvert F_p\rvert<F_1$ and follows the hysteresis cycle (1)-(4).}
    \label{fig:bistability_illustration}
\end{figure}

We begin by describing the relationship between the density of polaritons, $n$, and the intensity of the pump laser, $\lvert F_p\rvert^2$ in the case where the energy of the laser is above that of the lower polaritons, $\Delta_p > \gamma\sqrt{3}/2$: we square Eq.~\eqref{eq:homddGPE} and find
\begin{equation}
\left((gn - \Delta_p)^2 + \frac{\gamma^2}{4}\right) n = \lvert F_p\rvert^2
\end{equation}
or, equivalently,
\begin{equation}
        \left((\frac{m^* c_B^2}{\hbar} - \Delta_p)^2 + \frac{\gamma^2}{4}\right) \frac{m^* c_B^2}{g\hbar} = \lvert F_p\rvert ^2.
\end{equation}
The physics at play may be investigated equivalently in terms of the relationship between the speed of excitations $c_B$ and the strength of the pump, as shown in Figure~\ref{fig:bistability_illustration}.
At first, $c_B$ increases slowly with $\lvert F_p\rvert$ (arrow (1)), until $\lvert F_p\rvert=F_1$ where it increases abruptly (arrow (2)).
For $\lvert F_p\rvert>F_1$, $c_B$ increases slowly again.
If the pump's strength is decreased from $\lvert F_p\rvert\geq F_1$, $c_B$ decreases slowly until $\lvert F_p\rvert =F_2$ (arrow (3)), where it falls abruptly (arrow (4)).
Since $F_1>F_2$, the $c_B$ to $\lvert F_p\rvert$ relationship presents a hysteresis cycle with two regimes of speed of sound: the low density regime when $\lvert F_p\rvert<F_1$ and $c_B$ is low, and the high density regime when $\lvert F_p\rvert>F_2$ and $c_B$ is high.
This hysteresis cycle is the manifestation of optical bistability~\cite{baas_bista_2004}, so we will henceforth refer to it as the `bistability loop'.
Note that the dashed line in Fig.~\ref{fig:bistability_illustration} is unstable and the speed of sound will actually follow the hysteresis cycle schematised by arrows $(1)-(4)$.

\begin{figure}
    \centering
    \includegraphics[width= \columnwidth]{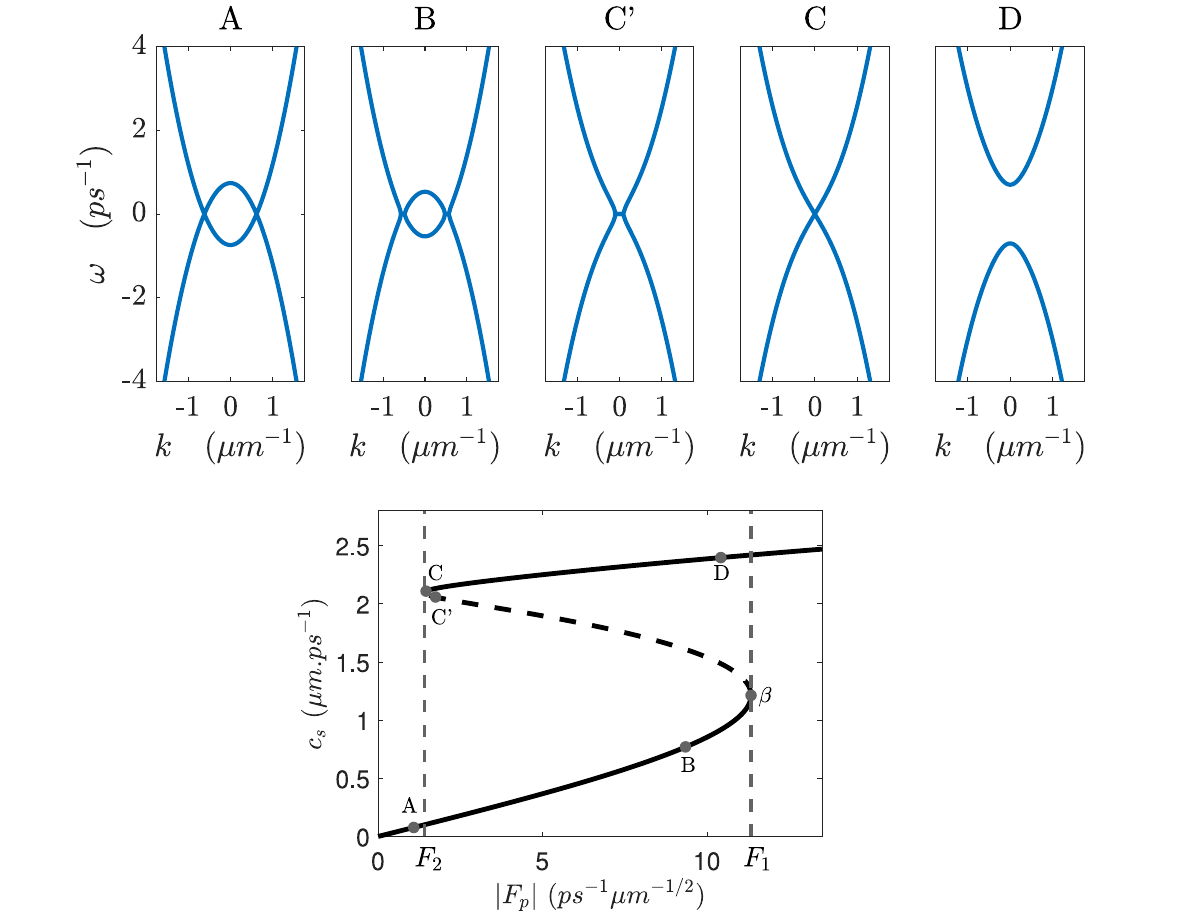}
    \caption{\textbf{Bogoliubov dispersion for various fluid densities.}
    Top row: Eq.\eqref{eq:bog_disp_polariton_fluid} is plotted in the fluid frame. \textbf{a)} and \textbf{b)}, regime of low density; \textbf{c)}, unstable fluid; \textbf{d)}, sonic dispersion; \textbf{e)} regime of high density.
    Bottom, \textbf{f)}, bistability curve for a homogeneous fluid.
    \textbf{A}, low density; \textbf{B}, low density and bistable; \textbf{C'}, unstable; \textbf{C}, sonic point; \textbf{D} high density and bistable.}
    \label{fig:bistability_disp_illu}
\end{figure}

Now, in order to explicitly show the dependence of the Bogoliubov dispersion on the density of the fluid as well as the influence of optical bistability thereon, we generalise Eq.~\eqref{eq:bog_disp_C}: We diagonalise the Bogoliubov matrix $\mathcal{L}$ for a homogeneous system pumped with arbitrary strength and obtain
\begin{equation}\label{eq:bog_disp_polariton_fluid}
    \omega_{\pm} (k) = \pm \sqrt{
                                    \left(
                                        \frac{\hbar k^2}{2 m^*} + 2gn - \Delta_p\right)^2
                                    - (gn)^2} - i \gamma/2.  
\end{equation}
Eq.~\eqref{eq:bog_disp_polariton_lab} is the Doppler shifted version of Eq.~\eqref{eq:bog_disp_polariton_fluid}.
In Fig.~\ref{fig:bistability_disp_illu}, we show the dispersion curve for 5 different fluid densities along the bistability loop.
As can be seen in Fig.~\ref{fig:bistability_disp_illu} \textbf{(a)} and \textbf{(b)}, the shape of the dispersion does not change much in the linear regime: the two branches of the dispersion curve cross.
When the fluid is bistable ($F_1\leq\lvert F_p\rvert\leq F_2$), in Fig.~\ref{fig:bistability_disp_illu} \textbf{(b)}, we observe the appearance of plateaus characteristic of an unstable fluid at the crossing points.
On the other hand, the shape of the dispersion curve changes significantly in the high density regime depending on the position along the bistability loop: at high pump strength (Fig.~\ref{fig:bistability_disp_illu} \textbf{(e)}), the two branches are split in energy by a gap that increases with the pump strength.
The sonic dispersion relation~\eqref{eq:bog_disp_C} is recovered at point $C$ (Fig.~\ref{fig:bistability_disp_illu} \textbf{(c)}), while for slightly lower pump strength (Fig.~\ref{fig:bistability_disp_illu} \textbf{(d)}), the plateau at low $k$ is characteristic of an unstable fluid (similarly to Fig.~\ref{fig:bistability_disp_illu} \textbf{(d)}) .
Note that the dispersion curve has a linear slope at low $k$ (and thus a sonic interpretation) at point $C$ only, which is thus sometimes referred to as the `sonic point' of the bistability.

As Eq.~\eqref{eq:bog_disp_polariton_fluid} is of order four in $k$, the dispersion has four complex roots.
The real part of these roots is non-zero in the low density regime (Fig.~\ref{fig:bistability_disp_illu} \textbf{(a)} and \textbf{(b)}) as well as at points $C$ and $C'$ (Fig.~\ref{fig:bistability_disp_illu} \textbf{(c)} and \textbf{d)}), but not at point $D$ (Fig.~\ref{fig:bistability_disp_illu} \textbf{(e)}).

In this appendix, we have seen that the mean-field of a polariton system behaves as a fluid.
We have reviewed the dispersion relation of Bogoliubov excitations in this fluid and seen that optical bistability of the fluid strongly influences the properties of this dispersion relation.
These considerations may be generalised to a fluid whose density is not homogeneous.

\section{Numerical method and correlation function}\label{app:twa}
In this appendix we present the numerical method used to compute the correlation maps of the main text.

Our interest is in vacuum emission, that is amplification of the quantum vacuum fluctuations at the horizon (the \textit{spontaneous} Hawking effect).
The quantum description of the Bogoliubov excitations relies on the dispersion relation of the classical field, Eq.~\eqref{eq:bog_disp_polariton_lab}.
In order to encompass quantum effects, we use a quantum Monte-Carlo method called the truncated Wigner approximation (TWA).
In this method, the equation of motion is truncated so as to map it to a stochastic partial differential equation for a classical field $\Psi$:
\begin{equation}\label{eq:stoGPE}
    i\dd \Psi = \left(\omega_0 - \frac{\hbar}{2m^*}\dv[2]{x} + V + g(\lvert \Psi\rvert^2 - 1/\Delta x) - i\frac{\gamma}{2}\right)\Psi\dd t+ F_p \dd t + \sqrt{\frac{\gamma}{4 \Delta x}} \dd W,
\end{equation}
where $\dd W$ is complex white noise.
In numerical simulations, sampling of the realisations obtained with~\eqref{eq:stoGPE} starts when the steady state is reached.
One must ensure that enough time is spent between each sampling to ensure independence of the realisations.
Quantum observables are computed with statistical averaging over the realisations obtained with the TWA: the general rule for $N$ arbitrary observables is \cite{carusotto_parametric_threshold_2005}
\begin{equation}\label{eq:twaobservables}
    \expval{O_1...O_N}_W
    = \frac{1}{N!} \sum_{\tiny{\substack{\text{All} \\\text{$N$-permutations}}}} \expval{P(\hat O_1,...,\hat O_N)},
\end{equation}
where $\expval{}_W$ denotes the statistical averaging over the realisations.
In all figures of this paper we computed 1 million realisations.

Emission at the horizon by the HE is best detected via nonlocal correlations in the fluid density~\cite{carusotto_numerical_2008}.
These may be quantified via the normalised spatial correlation function
\begin{equation}
\label{eq:g2app}
    g^{(2)}(x,x') = \frac{G^{(2)}(x,x')}{G^{(1)}(x)G^{(1)}(x')}.
\end{equation}
$G^{(2)}(x,x')$ is the diagonal two-points correlation function of the field, which is calculated from~\eqref{eq:twaobservables} and normally ordered using Bose statistics
\begin{equation}
\begin{split}
\label{eq:bosestatistics}
    G^{(2)}(x,x') & = \expval{\hat \Psi^\dagger(x) \hat \Psi^\dagger(x') \hat \Psi(x') \hat \Psi (x)}\\
                  & = \expval{ \Psi^* (x) \Psi^* (x') \Psi (x') \Psi (x) }_W -\frac{1}{2\Delta x} (1+\delta_{x,x'})     \times\\
                  &\,\,\,\,\,  \bigg(\expval{ \Psi^* (x) \Psi (x) }_W +\expval{ \Psi^* (x') \Psi (x') }_W -\frac{1}{2 \Delta x} \bigg),
\end{split}
\end{equation}
while the diagonal one-point correlation function is
\begin{equation}
    G^{(1)}(x) = \expval{\hat\Psi^\dagger (x) \hat\Psi(x)} = \expval{\Psi^*(x)\Psi(x)}_W - \frac{1}{2 \Delta x}.
\end{equation}

\section{Constraints on the calculations}\label{app:constraints}

All configurations in Table.~\ref{fig:corr} have been realised with the cavity parameters of~\cite{Nguyen}.
When exploring all possible configurations of fluid density on either side of the horizon, some constraints must be abode by.
We present them in this appendix.

The first constraint is on the upstream pump wavevector $k_{p,u}$ for a fluid near the sonic point.
The fluid is at the sonic point for
\begin{equation}
    c_u = \sqrt{\frac{\omega_p-\omega_0-\hbar k_{p,u}^2/2m^*}{m^*}},
\end{equation}
together with the upstream condition $v_u<c_u$, this yields an upper bound for the upstream fluid flow velocity and thus for the wavevector of the pump:
\begin{equation}
    \omega_p-\omega_0>\frac{3}{2}m^* v_u^2.
\end{equation}
For the value of detuning used throughout this paper the upper bound is around $k_{p,u} = \SI{0.28}{\per\micro\meter}$. In most simulations, we used $k_{p,u} = \SI{0.25}{\per\micro\meter}$ in order to be close to the bound while leaving a small interval for easier simulations.

Exploring all regimes of density in the downstream region comes with some constraints as well: For instance, placing the fluid in the upper part of the bistable regime as in configurations~\ref{fig:corr} \textbf{(b)} is easier for a large bistable interval $F_1-F_2$. Point $\beta$ in configuration~\ref{fig:bistability_illustration} \textbf{f)} is obtained at
\begin{equation}
    c_d = \sqrt{\frac{\omega_p - \omega_0 - \hbar k_{p,d}^2/2m^*}{2m^*}}.
\end{equation}
and the width of the interval is then given by
\begin{equation}
\begin{split}
    \Delta I^{bistable}
    & = \lvert F_{p,{max}}\rvert^2 - \lvert F_{p,{min}}\rvert^2 \\
    & = \left(
             \frac{4}{9}\left(\omega_p-\omega_0-\frac{\hbar k_{p,d}^2}{2m^*}\right)^2 - \frac{1}{2}\gamma^2\right)\\
        &\,\,\,\,\,\times\frac{\omega_p-\omega_0-\hbar k_{p,d}^2/2m^*}{3g}.
\end{split}
\end{equation}
A bistable regime exists only if $\omega_p-\omega_0>\hbar k_{p,d}^2/2m^*+\gamma\nicefrac{\sqrt{3}}{2}$, hence an upper bound on $k_{p,d}$.

Furthermore, the speed of excitations right after the defect, $c_{d}$ is fixed by the upstream parameters and the strength of the defect, $V_{def}$. Pumping in the upper branch of the bistability requires
\begin{equation}
    c_d > \sqrt{\frac{\omega_p-\omega_0 - \hbar k_{p,d}^2/2m^*}{m^*}}.
\end{equation}
This critical point needs to be below $c_{d}$ for the fluid density to be on the upper branch.
(Note that it would also be possible to change the value of $c_{d}$, which can be achieved for a weaker defect potential --- energy conservation before and just after the defect links $V_{def}$ and $c_{d}$).
The choice of different $k_{p,d}$ in the simulations of Fig.~\ref{fig:corr} is a consequence of all these constraints.

\section{Repulsive defect}\label{app:repdef}
Here we consider the configuration of Fig.~\ref{fig:corr}~\textbf{(e)} but with a repulsive defect ($V_{ext}=0.85meV$) instead.

\begin{figure}
    \centering
    \includegraphics[width=\columnwidth]{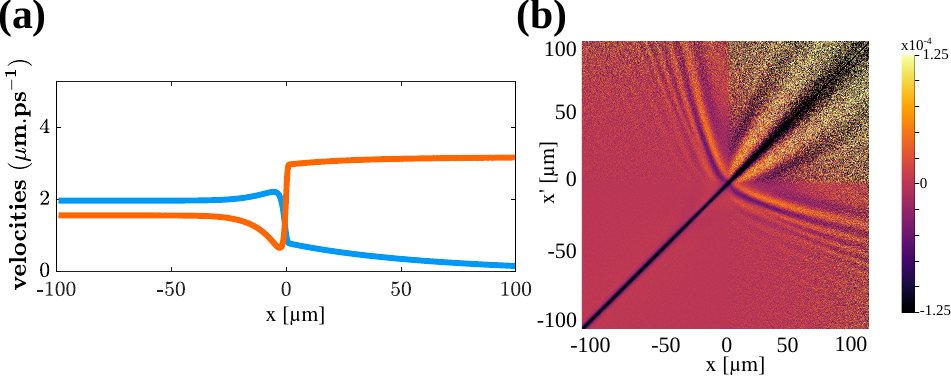}
    \caption{\textbf{Emission with a repulsive defect.} \textbf{(a)} Flow profile. \textbf{(b)} Correlated emission.}
    \label{fig:repdefect}
\end{figure}

This configuration is comparable to that of ref~\cite{Gerace} but with a pumping scheme such that the fluid density is supported as close to point $C$ of the bistability as possible.
In Fig.~\ref{fig:repdefect}, we observe that, although stronger and longer than in~\cite{Gerace}, correlated emission by the Hawking effect is weaker than in Fig.~\ref{fig:corr}~\textbf{(e)}.
So the attractive potential of the defect aids correlated emission.
Furthermore, the horizontal/vertical traces signalling correlations with the horizon region (vacuum quantum excitation of a QNM of the acoustic field~\cite{jolyInterplayHawkingEffect2021}) are absent.

\end{appendices}


\end{document}